\documentclass[aps,prd,showpacs,twocolumn]{revtex4}

\usepackage{epsfig}
\usepackage{longtable}

\begin{document}

\title{The  $D_{sJ}^*(2860)$ Mesons as  Excited D-wave $c\bar{s}$ States} 
\author{ Stephen Godfrey\footnote{Email: godfrey@physics.carleton.ca} and Kenneth Moats}
\affiliation{
Ottawa-Carleton Institute for Physics, 
Department of Physics, Carleton University, Ottawa, Canada K1S 5B6 
}

\date{November 4, 2015}

\begin{abstract}
A new charm-strange meson,  the $D_{s1}^*(2863)$, has recently been observed by the LHCb collaboration 
which also determined the $D_{sJ}^*(2860)$ to have spin 3.  One of
the speculations about the previously observed $D_{s1}^*(2710)$ is that it is the
$1^3D_1(c\bar{s})$ state. In this paper we reexamine the quark model properties and assignments 
of these three states in light of these new measurements. 
We conclude that the $D_{s1}^*(2863)$ and $D_{s3}^*(2863)$ are the $1^3D_1 (c\bar{s})$ 
and $1^3D_3 (c\bar{s})$ states respectively and the $D_{s1}^*(2710)$ is the $2^3S_1(c\bar{s})$ state. 
In addition to these three states there are another three excited $D_s$ states in this mass region 
still to be found; the $2^1S_0$ and two $1D_2$ states.
We calculate the properties of these states and expect that LHCb has the capability of
observing these states in the near future. 
\end{abstract}

\maketitle

\section{Introduction}

Over the past few years a proliferation 
 of new meson states have been observed by various 
collider experiments leading to a renaissance 
in hadron spectroscopy 
\cite{Swanson:2006st,Godfrey:2008nc,Godfrey:2009qe,Braaten:2013oba,DeFazio:2012sg}.  
While many of these new states do not fit into the quark model description of hadrons 
and remain enigmas, some, on the surface, appear to be conventional quark model states. 
The challenge is then of classification and being able to describe the properties of the 
latter set as conventional quark model states.  Sharpening our descriptive power for 
conventional states is a necessary prerequisite to understanding the nature of the exotic states. 

Recently the LHCb collaboration presented evidence for overlapping spin-1 and spin-3 
$\bar{D}^0K^-$ resonances at 2.86 GeV/c$^2$ \cite{Aaij:2014xza,Aaij:2014baa}.  
These follow the observation of three new excited charm-strange mesons:  
the $D_{s1}^*(2700)^\pm$ \cite{Aubert:2006mh,Brodzicka:2007aa,Aubert:2009ah,Aaij:2012pc},
$D_{sJ}^*(2860)^\pm$ \cite{Aubert:2006mh,Aubert:2009ah,Aaij:2012pc}, 
and $D_{sJ}^*(3040)^+$ \cite{Aubert:2009ah}. It is the first two states that are 
most relevant to the new LHCb measurements.  

The $D_{s1}^*(2710)^\pm$
has been identified with the first radial excitation of the $D_{s1}^*(2112)^\pm$ 
or the  $D_{s}^*(1^3D_1)$ or some mixture of them 
\cite{Close:2006gr,Zhang:2006yj,Colangelo:2007ds,Zhong:2009sk,Zhang:2009nu,Chen:2009zt,Li:2009qu,Ebert:2009ua,Li:2010vx,Chen:2011rr,Wang:2013mml,Yuan:2012ej,Godfrey:2013aaa,Colangelo:2012xi} 
and the $D_{sJ}^*(2860)^\pm$ as the $D_{s}^*(1^3D_1)$ or the  $D_{s}^*(1^3D_3)$ 
\cite{Colangelo:2006rq,Zhang:2006yj,Zhong:2009sk,Zhang:2009nu,Chen:2009zt,Li:2009qu,Li:2010vx,Chen:2011rr,Yuan:2012ej,Godfrey:2013aaa,Colangelo:2012xi}.  
The theoretical predictions for these states are not totally consistent
with their observed properties and it was suggested that a quark model identification that best described 
the observed properties is to identify $D_{s1}^*(2710)^\pm$ as the 
$D_{s}^*(1^3D_1)$ and the $D_{sJ}^*(2860)^\pm$ as the $D_{s}^*(1^3D_3)$ but overlapping with the 
$1D_2(c\bar{s})$ states to explain the observed $D^*K/DK$ branching ratios \cite{Godfrey:2013aaa,Zhong:2009sk,Colangelo:2012xi}.  
Furthermore, it was proposed that studying the angular distributions of the final states would test this possibility.  
LHCb did this analysis and found that the $D_{sJ}^*$ 
was actually comprised of $J=1$ and $J=3$ states. The LHCb measurements 
undermine the explanation given above and suggest 
that a reexamination of these states is warranted.

The LHCb collaboration determined the masses and widths of the $D_{s1}^*(2860)$ and $D_{s3}^*(2860)$ to be \cite{Aaij:2014xza,Aaij:2014baa}:
\begin{equation}
M(D^*_{s1}(2860)^- ) = 2859 \pm 12 \pm 6 \pm 23 \; \hbox{MeV} 
\end{equation}
\begin{equation}
\Gamma(D^*_{s1}(2860)^- ) = 159 \pm 23 \pm 27 \pm 72 \; \hbox{MeV}
\end{equation}
\begin{equation}
M(D^*_{s3}(2860)^- ) = 2860.5 \pm 2.6 \pm 2.5 \pm 6.0 \; \hbox{MeV} 
\end{equation}
\begin{equation}
\Gamma(D^*_{s3}(2860)^- ) = 53 \pm 7 \pm 4 \pm 6  \; \hbox{MeV}
\end{equation}
where the first uncertainty is statistical, the second is due to experimental systematic effects and the third is due to model variations.  
The Particle Data Group averages for the masses, decay widths and ratios of branching 
fractions for the $D_{s1}^*(2700)^\pm$ and $D_{sJ}^*(2860)^\pm$ 
are \cite{Beringer:1900zz}:
\begin{equation}
M(D^*_{s1}(2710)^\pm ) = 2709\pm 4 \; \hbox{MeV} 
\end{equation}
\begin{equation}
\Gamma(D^*_{s1}(2710)^\pm ) = 117 \pm 13 \; \hbox{MeV}
\end{equation}
\begin{equation}
{{\Gamma (D^*_{s1} \to D^* K)} \over {\Gamma (D^*_{s1} \to D K)}} 
= 0.91 \pm 0.13 (\hbox{stat}) \pm 0.12 (\hbox{syst})
\end{equation}
and
\begin{equation}
M(D^*_{sJ}(2860)^\pm ) = 2863^{+4.0}_{-2.6} \; \hbox{MeV} 
\end{equation}
\begin{equation}
\Gamma(D^*_{sJ}(2860)^\pm ) = 58 \pm 11 \; \hbox{MeV}
\end{equation}
\begin{equation}
{{\Gamma (D^*_{sJ} \to D^* K)} \over {\Gamma (D^*_{sJ} \to D K)}} 
= 1.10\pm 0.15 (\hbox{stat}) \pm 0.19 (\hbox{syst}).
\end{equation}
Both states are observed decaying into both $DK$ and $D^*K$ so have natural
parity $J^P= 1^-$, $2^+$, $3^-, \ldots$. 

In this paper we reexamine these states in light of the new LHCb measurements by comparing the observed properties to the mass predictions of the relativized quark model \cite{godfrey85xj}
and the decay predictions of the $^3P_0$ pair creation decay model \cite{Micu:1968mk,Le Yaouanc:1972ae,Ackleh:1996yt,Blundell:1995ev}.  In the following section we compare the observed masses to the predictions of the relativized quark model for charm-strange mesons \cite{godfrey85xj,Godfrey:2013aaa} and give the partial decay widths for the $2S$ and $1D$ 
charm-strange mesons calculated using the $^3P_0$ model.  In section III we discuss these results and we give a brief summary in section IV. 

\begin{table}[h]
\caption{Partial widths for the $2S$ and $1D$ $c\bar{s}$ mesons calculated using the 
$^3P_0$ quark pair creation model.
The widths in column 4 were calculated using the predicted masses for the initial states 
and the PDG values \cite{Beringer:1900zz} for the final states.
The widths given in the last column were calculated using $M= 2709$~MeV for
the $2^3S_1$ initial state and $M= 2859$ and 2863~MeV for the $1^3D_1$ and $1^3D_3$ initial states
respectively. The $2^1S_0$ and $D_2$ masses were obtained by subtracting the predicted splittings 
using the masses given in column 4 
from the measured $2^3S_1$ and $1^3D_3$ masses as described in the text.
\label{tab:properties}}
\begin{tabular}{llccc} \hline \hline
State & Property  & Experiment & \multicolumn{2}{c}{Predicted} \\
	&	& (MeV) & \multicolumn{2}{c}{(MeV)}  \\
\hline 
$D_s^*(2^3S_1)$ 	& Mass 							& $2709\pm 4$ 		& 2732 	& 2709\footnote{input as described in the text} 	\\
				& $D_s^*(2^3S_1)\to DK$ 			& 					& 41.      	& 40.			\\	
				& $D_s^*(2^3S_1)\to D^*K$ 			& 					& 82.		& 75.			\\	
				& $D_s^*(2^3S_1)\to D_s\eta$ 		& 					& 7.9		& 6.9			\\	
				& $D_s^*(2^3S_1)\to D_s^*\eta$ 		& 					& 5.4		& 3.1			\\	
    			& $\Gamma_{\hbox{Total}}$ 			& $117\pm 13$ 			& 136.	& 125.		\\	
    			& $\Gamma(\to D^* K)/\Gamma(\to DK)$ 	& $0.91 \pm 0.18$  			& 2.0		& 1.8			\\	
\hline
$D_s(2^1S_0)$ 	& Mass 							& 					& 2673 	& 2650$^a$ 	\\
				& $D_s(2^1S_0)\to D^*K$ 			& 					& 94.		& 78.			\\	
				& $D_s(2^1S_0)\to D_s^*\eta$ 		& 					& 0.6		& 0			\\	
				& $\Gamma_{\hbox{Total}}$ 			&  					& 94.		& 78.			\\	
\hline
$D_s^*(1^3D_3)$ 	& Mass 							& $2863^{+4.0}_{-2.6}$ 	& 2917 	& 2863$^a$ 	\\
				& $D_s^*(1^3D_3)\to DK$ 			&  					& 28.		& 20.			\\	
				& $D_s^*(1^3D_3)\to D^*K$ 			& 					& 20.		& 12.			\\	
				& $D_s^*(1^3D_3)\to DK^*$ 			&  					& 1.9		& 0.4			\\	
				& $D_s^*(1^3D_3)\to D^*K^*$ 			&  					& 12.		& 0			\\	
				& $D_s^*(1^3D_3)\to D_s\eta$ 		&  					& 1.7		& 1.0			\\	
				& $D_s^*(1^3D_3)\to D_s^*\eta$ 		&  					& 0.7		& 0.3			\\	
   				& $\Gamma_{\hbox{Total}}$ 			& $58 \pm 11$ 			& 64.		& 34.			\\	
    			& $\Gamma(\to D^* K)/\Gamma(\to DK)$ 	& $0.91 \pm 0.18$  			& 0.72	& 0.62		\\	
\hline
$D_s(D_{2}')$ 		& Mass 							& 					& 2926 	& 2872$^a$	\\	
				& ${D_s}_{2}'\to D^*K$ 				&  					& 163	& 159		\\	
				& ${D_s}_{2}'\to DK^*$ 				&  					& 8.7		& 4.4			\\	
				& ${D_s}_{2}'\to D^*K^*$ 				&  					& 7.9		& 0			\\	
				& ${D_s}_{2}'\to D_s^*\eta$ 			& 					& 26		& 21			\\	
  				& $\Gamma_{\hbox{Total}}$ 			&  					& 206	& 184		\\	
\hline
$D_s(D_{2})$ 		& Mass 							& 					& 2900 	& 2846$^a$	\\	
				& ${D_s}_{2}\to D^*K$ 				& 					& 28		& 16			\\	
				& ${D_s}_{2}\to DK^*$ 				&  					& 101	& 58			\\	
				& ${D_s}_{2}\to D^*K^*$ 				&  					& 0.07	& 0			\\	
				& ${D_s}_{2}\to D_s^*\eta$ 			& 					& 0.9		& 0.4			\\	
   				& $\Gamma_{\hbox{Total}}$ 			&  					& 130.	& 75			\\	
\hline
$D_s^*(1^3D_1)$ 	& Mass 							& $2859 \pm 27$ 		& 2899 	& 2859$^a$ 	\\	
				& $D_s^*(1^3D_1)\to DK$ 			& 					& 93.		& 94.			\\	
				& $D_s^*(1^3D_1)\to D^*K$ 			&  					& 51.		& 49.			\\	
				& $D_s^*(1^3D_1)\to DK^*$ 			&  					& 32.		& 22.			\\	
				& $D_s^*(1^3D_1)\to D_s\eta$ 		&  					& 18.		& 16.			\\	
				& $D_s^*(1^3D_1)\to D_s^*\eta$ 		& 					& 7.2		& 5.8			\\	
   				& $\Gamma_{\hbox{Total}}$ 			& $159\pm 80$ 		& 201.	& 187.		\\	
   				& $\Gamma(\to D^* K)/\Gamma(\to DK)$ 	&  					& 0.55	& 0.52		\\	
\hline
\hline
\end{tabular}

\end{table}

\section{$2S$ and $1D$ $D_s$ Properties}

\subsection{Spectroscopy}

We compare the observed masses to the predictions for the charm-strange mesons of the 
relativized quark model \cite{Godfrey:2013aaa}
in Table~\ref{tab:properties}.  The details of this model can 
be found in Ref.~\cite{godfrey85xj} 
and \cite{Godfrey:1985by,godfrey85b,Godfrey:1986wj,Godfrey:2004ya,Godfrey:2005ww} to 
which we refer the interested reader. The parameters of the model, including the 
constituent quark masses, are given in Ref.~\cite{godfrey85xj}.
This model has been reasonably successful in 
describing most known mesons although in recent years an increasing number 
of states 
have been observed that do not fit into this picture and are often referred to 
as ``exotics'' \cite{Swanson:2006st,Godfrey:2008nc,Godfrey:2009qe,Braaten:2013oba,DeFazio:2012sg}. 
An important limitation of this model is that it is restricted to the $q\bar{q}$ 
sector of the Fock space and does not take into account higher components that 
can be described by coupled channel effects 
\cite{Eichten:2004uh,Barnes:2007xu,Geiger:1992va}.  As a consequence 
of neglecting these effects
and the crudeness of the relativization procedure we do not 
expect the mass predictions to be accurate to better than $\sim 10-20$~MeV.

For the case of a quark and antiquark of unequal mass, charge conjugation
parity is no longer a good quantum number so that states with different 
total spins but with the same total angular momentum, such as
$^3P_1 -^1P_1$ and $^3D_2 -^1D_2$ pairs, can mix via
the spin orbit interaction or some other mechanism such as mixing via coupled channels.
Consequently, the physical $J=2$ $D$-wave states are linear
combinations of $^3D_2$ and $^1D_2$ which we describe by:
\begin{eqnarray}
\label{eqn:mixing}
D_{2}  & = {^1D_2} \cos\theta_{nD} + {^3D_2} \sin\theta_{nD} \nonumber \\   
D_{2}^{\prime} & =-{^1D_2} \sin\theta_{nD} + {^3D_2} \cos \theta_{nD} 
\end{eqnarray}
where $D\equiv L=2$ designates the relative orbital angular momentum of the $q\bar{q}$ 
pair and the subscript $J=2$ is the total angular momentum including spin of the $q\bar{q}$ 
pair which is equal to $L$ with analogous expressions for other values of $L$.  
This notation implicitly implies $L-S$ 
coupling between the quark spins and the relative orbital angular momentum.  
$\theta_{1D}$ is found by diagonalizing the mass matrix for the antisymmetric
piece of the spin-orbit interaction (which arises for unequal mass quarks and antiquarks)
in the basis of eigenvectors of the $|jm; ls\rangle$ sectors. 
We obtain $\theta_{1D}=-38.5^\circ$ (for $c\bar{s}$) \cite{Godfrey:2013aaa}.  
The details are given in Ref.~\cite{godfrey85xj}.
In the heavy quark limit (HQL) in which the heavy quark mass $m_Q\to \infty$, 
the states can be described by the total angular momentum of the
light quark, $j_q$, which couples to the spin of the heavy quark and
corresponds to $j-j$ coupling. In this limit the state that is mainly spin singlet has $j_q=l+{1\over 2}$
while the state that is mainly spin triplet has $j_q=l-{1\over 2}$ and is labelled with a prime \cite{Cahn:2003cw}.
For $L=2$ the HQL gives 
rise to two doublets,  $j_q=3/2$ and $j_q=5/2$ with $\theta_{D}=-\tan^{-1}(\sqrt{2/3})=-39.2^\circ$ 
where the minus sign arises from our $c\bar{s}$ convention
\cite{Godfrey:1986wj,Cahn:2003cw,Zhang:2009nu,DiPierro:2001uu,Colangelo:2006rq,Colangelo:2007ds,Chen:2009zt,Chen:2011rr}.  
We note that the 
definition of the mixing angles is fraught 
with ambiguities and
one should be extremely careful comparing predictions from different 
papers \cite{barnes}.

\subsection{Strong Decays}

We calculate decay widths using the $^3P_0$ quark creation model 
\cite{Micu:1968mk,Le Yaouanc:1972ae,Ackleh:1996yt,Blundell:1995ev,Barnes:2005pb}.
There are a number of predictions for $D_s$ decay widths in the literature using the $^3P_0$ model 
\cite{Close:2005se,Close:2006gr,Zhang:2006yj,Li:2009qu,Li:2010vx,Yuan:2012ej}
and other models \cite{Colangelo:2006rq,Colangelo:2007ds,Zhong:2009sk,DiPierro:2001uu,Chen:2011rr}.

In our calculations we use for the quark creation parameter $\gamma=0.4$  which has been 
found to give a good description of strong decays \cite{Barnes:2005pb,Close:2005se}.
We use harmonic oscillator wave functions with the oscillator parameter, $\beta_{c\bar{s}}$,
obtained by equating the rms radius of the harmonic oscillator wavefunction for the 
specified $(n,l)$ quantum numbers to the 
rms radius of the wavefunctions calculated using the 
relativized quark model of Ref.~\cite{godfrey85xj} except for the light mesons 
for which we use a universal $\beta =0.40$~GeV (see also Ref.~\cite{Close:2005se,Blundell:1995ev}).  
The harmonic oscillator wavefunction parameters found in this way
 are:  $\beta_{c\bar{s}}(2^3S_1)=0.46$~GeV, 
$\beta_{c\bar{s}}(2^1S_0)=0.48$~GeV, $\beta_{c\bar{s}}(1^3D_3)=0.43$~GeV, 
$\beta_{c\bar{s}}(1^3D_2)=0.45$~GeV, $\beta_{c\bar{s}}(1^1D_2)=0.44$~GeV, 
$\beta_{c\bar{s}}(1^3D_1)=0.47$~GeV $\beta_{c\bar{s}}(1^3S_1)=0.56$~GeV, 
$\beta_{c\bar{s}}(1^1S_0)=0.65$~GeV, $\beta_{c\bar{q}}(1^3S_1)=0.52$~GeV, 
and $\beta_{c\bar{q}}(1^1S_0)=0.60$~GeV.  For the constituent 
quark masses in 
our calculations
of both the meson masses and of the strong decay widths 
we use
$m_c=1.628$~GeV, $m_s=0.419$~GeV and $m_q=0.220$~GeV.  
Finally, we use ``relativistic phase space'' as described in Ref.~\cite{Blundell:1995ev,Ackleh:1996yt}.

As stated above we used standard values of $\gamma=0.4$ and $\beta=0.4$~GeV for 
the light quark mesons.  These typical values were found from fits to light meson
decays \cite{Close:2005se,Blundell:1995ev,Blundell:1996as}. The predicted widths
are fairly insensitive to the precise values used for $\beta$ provided $\gamma$ 
is appropriately rescaled.  However $\gamma$ can vary as much as 30\% and still
give reasonable overall fits of light meson decay widths \cite{Close:2005se,Blundell:1996as}.
This can result in factor of two changes to predicted widths, both smaller or larger.

The resulting partial widths for the $2S$ and $1D$ multiplets are given in Table~\ref{tab:properties}. 
The widths given in column 4 were obtained using the predicted masses for the excited 
states and the PDG values for the decay products while the widths given in column 5 use 
the measured masses for the $D_{s1}^*$ and $D_{s3}^*$ states as input values: 
$M(2^3S_1)=M(D_{s1}^*)=2709$~MeV, $M(1^3D_1)=M(D_{s1}^*)=2859$~MeV and $M(1^3D_3)=M(D_{sJ}^*)=2863$~MeV. 
We also use these observed masses to shift our input masses for the remaining states.
For the $2^1S_0$ mass we subtracted the predicted $2^3S_1 - 2^1S_0$ splitting from
the measured $D_{s1}^*(2710)$ mass and for the $D_2$ states we calculated 
the predicted mass differences
with respect to the $D_{s3}^*$ state  and subtracted them from 
the observed $D_{sJ}^*(2860)$ mass.  In all cases we calculated the mass splittings
using the mass predictions given in column 4.

\section{Discussion}

The predicted masses 
and widths of the $D_s^*(2^3S_1)$,  $D_s^*(1^3D_1)$ and $D_s^*(1^3D_3)$  are in reasonably good 
agreement with the observed properties of the $D_{s1}^*(2710)^\pm$, 
$D_{s1}^*(2860)$ and $D_{s3}^*(2860)$. The largest discrepancy in the masses is for the $1^3D_3$ state.  The predicted 
total widths agree with the measured widths within the experimental error and the expected predictive power of the decay model.  
In addition, the predicted value for the ratio of branching ratios  ${\cal B}(1^3D_3 \to D^*K)/{\cal B}(1^3D_3 \to DK)$ 
is in reasonable agreement with the measured ratio but the analogous ratio for the $2^3S_1$ state is roughly a factor of 
two larger than the measured value.  This is the largest discrepancy between predicted and observed quantities.

As suggested in previous studies the $D_{s1}^*(2710)^\pm$ properties could be explained
by treating it as a mixture of $2^3S_1(c\bar{s})$ and $1^3D_1(c\bar{s})$ 
\cite{Godfrey:2013aaa,Close:2006gr,Zhong:2009sk,Chen:2011rr,Li:2010vx,Li:2009qu,Yuan:2012ej,Wang:2013mml}. 
We find that a small $2^3S_1 - 1^3D_1$ mixing angle   
 of $\sim 10^\circ$ and $7.3^\circ \; (5.1^\circ)$ brings the   
${\cal B}(2^3S_1 \to D^*K)/{\cal B}(2^3S_1 \to DK)$ ratio to the central value and within one (two) 
standard deviations of the measured value. 
With these mixings  we find that for the orthogonal partner, the $D_{s1}^*(2860)$, the 
${\cal B}(1^3D_1 \to D^*K)/{\cal B}(1^3D_1 \to DK)$ ratio is 
1.04, 0.87 and 0.75 
for $\theta = 10^\circ$, 
$7.3^\circ$ and $5.1^\circ$ respectively.
Thus, measuring the $D^*K$ branching ratio of the $D_{s1}^*(2860)$ 
would be a consistency check for this description of the $D_{s1}^*(2710)^\pm$ meson.

Overall 
the three recently discovered excited charm-strange mesons are well 
described as the $2^3S_1$, $1^3D_1$ and $1^3D_3$ charm-strange mesons. 
A recent paper by Song {\it et al} \cite{Song:2014mha} also using the $^3P_0$ model but with different
input parameters gives results
in reasonable agreement to those reported here and comes to the same conclusions. 
However, many previous studies concluded that the $D_{s1}^*(2710)^\pm$ properties were inconsistent 
with that of the $D_s^*(2^3S_1)$ 
because the $D_s^*(2^3S_1)$ was predicted to be significantly narrower than the measured width 
\cite{Godfrey:2013aaa,Zhang:2006yj,Zhong:2009sk,Yuan:2012ej,Wang:2013mml,Li:2009qu,Li:2010vx} 
and the predicted ${\cal B}(2^3S_1 \to D^*K)/{\cal B}(2^3S_1 \to DK)$ ratio is much larger than what was measured.  
Although there were some exceptions to this conclusion \cite{Yuan:2012ej,Close:2006gr} the general consensus 
was that the $D_{s1}^*(2710)^\pm$ is the 
$D_s^*(1^3D_1)$ or perhaps a $2^3S_1 - 1^3D_1$ mixture \cite{Close:2006gr,Zhong:2009sk,Yuan:2012ej,Chen:2011rr,Li:2010vx}. 
So although the results we give in this paper are in good agreement with the observed properties, 
one should take the variation in predictions by different calculations 
as a cautionary reminder about the precision of 
the predictions.

Three of the six excited $c\bar{s}$ states 
in this mass region have now been identified.  We expect the spin-singlet partner 
of the $D_s^*(2710)$ (the $2^3S_1$) to  lie $\sim 60$~MeV lower in mass \cite{Godfrey:2013aaa}, 
$M(2^1S_0(c\bar{s}))\sim 2650$~MeV 
and predict its width to be $\sim 78$~MeV and decaying to $D^*K$. 
The  $2^3S_1 - 2^1S_0$ mass splitting was obtained using the predicted masses given in
Table~\ref{tab:properties}.
Similarly, taking the $1^3D_3$ $c\bar{s}$
mass to be that of the $D_{s3}^*(2860)$ and using the splittings between the $j=2$ states and 
the $1^3D_3$ state obtained from the predicted masses given in column 4 of 
Table~\ref{tab:properties} we expect the $D_2'$ mass to be $\sim 2872$~MeV  
and the $D_2$ to be $\sim 2846$~MeV.  Using these masses we obtain the
decay properties of the $j=2$ members of the D-wave multiplet given in column 5 of
Table~\ref{tab:properties}. 
In the heavy quark limit  one of the $j=2$ states is expected
to be degenerate with the $1^3D_3$ state and be relatively narrow while the other $j=2$ state
is expected to be degenerate with the $1^3D_1$ state and be relatively broad.  
A distinguishing feature of these states is that the broad $D_2'$ state is expected
to decay predominantly to $D^*K$ with a sizeable BR to $D^*_s \eta$ 
while the narrow $D_2$ state is expected to decay predominantly to
$DK^*$ with a sizeable BR to $D^*K$.   
We expect that with the expected increased statistics LHCb should 
be able to complete the $2S$ and $1D$ multiplets and further test and improve our understanding 
of this spectroscopy.

\section{Summary}

The observation of the $D_{s1}^*(2710)^\pm$  and $D_{sJ}^*(2860)^\pm$ mesons by the
BaBar \cite{Aubert:2006mh,Aubert:2009ah} and Belle collaborations \cite{Brodzicka:2007aa}
resulted in considerable theoretical interest which led to a number of possible quark
model assignments.  The recent LHCb results showing that there were in fact two
overlapping resonances at 2860~MeV \cite{Aaij:2014xza,Aaij:2014baa} with $j=1$ and $j=3$ added
considerably to our understanding.  The calculations presented in this paper indicate 
that the two states at 2860~MeV are almost certainly the $1^3D_1$ and $1^3D_3$ $c\bar{s}$ states
and the $D_{s1}^*(2710)^\pm$ is almost certainly the $2^3S_1 (c\bar{s})$ state.  The 
discrepancy between the predicted and measured $D_{s1}^*(2710)^\pm$ branching ratios to 
$D^*K$ and $DK$ can be accommodated with a small $2^3S_1-1^3D_1$ mixing.  Six excited
$D_s$ mesons are expected in this mass region leaving three of them still undiscovered.  
We predict the properties for these states and expect that the discovery of the missing states 
is well within 
the capabilities of LHCb in the near future.  Their observation and measurement 
of their properties would be a valuable test of the various models describing these states.

\noindent
{\bf Note added:}  At the completion of this paper, a paper by Wang appeared \cite{Wang:2014jua} which
analyzed the $D^*_{s1,3}$ states using a heavy meson effective Lagrangian with chiral 
symmetry breaking corrections.  He found that he could describe the  $D_{s3}^*(2860)$ branching ratios by 
suitable choice of model parameters but did not give numerical values for the partial widths.

\acknowledgments

This research was supported in part 
the Natural Sciences and Engineering Research Council of Canada under grant number 121209-2009 SAPIN. 
SG thanks the University of Toronto Physics Department for their hospitality where this 
work was completed.


\end{document}